\documentclass[11pt,twocolumn]{amsart}
\usepackage{geometry}                
\usepackage{graphicx}
\usepackage{amssymb}
\usepackage{epstopdf}
\title{Discovery \&\ Depth}
\author{S. R. Kulkarni}

\begin{document}
\onecolumn

\maketitle

\section{Background}
\label{sec:Background}

In the United States, the National Science Foundation (NSF) has
commissioned a review of NSF-funded astronomy assets with the goal
of determining how to best allocate funding for this decade.\footnote{The
introduction is to give some background for those readers of this
article who are not familiar with the US astronomy scene. This paper
sans \S\ref{sec:Recommendations} was submitted to PR2012 in response
to the requests for inputs from members of the US astronomical
community.} This ``Portfolio Review'' (hereafter PR2012) was motivated
by the perceived flat funding for Astronomy  for the rest of the
decade (from NSF) against the backdrop of funding annual operating
costs of current facilities, the anticipated  large burden of
operation costs for recent and future ``flagship'' facilities (ALMA,
LSST, GSMT), the desire to  maintain healthy line of ``mid-scale''
projects (innovative and focused; typically built and managed by
groups) and support individual researchers (``PI'' grants).  The
various financial outlays are reasonably understood. Thus the
principal task of PR2012 is the prioritization of the astronomical returns
from the assets discussed above.

Here, accepting the boundary conditions posed above, I have focused
on fields centered on optical astronomy which offers the best
opportunity for progress in this decade and thus offer the highest cost-benefit ratio. 
I have made an effort to
keep his analysis  separate from the personal recommendations
(\S\ref{sec:Recommendations}).  Readers may profit from reading
\S\ref{sec:AstronomyPhysics}--\S\ref{sec:OptimalBalance}.  Some
colleagues have noted that my personal recommendations do not
adhere to the rules of PR2012 (although the arguments are based on
objective reality).  Readers who are delicate or who are sticklers of rules and regulations
may find \S\ref{sec:Recommendations} stressful and are thus advised
to skip the same.

\section{Astronomy \&\ Physics}
\label{sec:AstronomyPhysics}

Astronomy, like Biology, is primarily a phenomenologically driven
subject. In contrast, Physics is a reductionist subject. A single
profound realization (e.g. Newton's formulation of gravity) can
keep a large army of physicists busy for long periods with figuring
out the ramification of this one realization.  Thus in physics deep
understanding is what drives the field. In contrast, in Astronomy
both \textit{Discovery} and \textit{Depth} are equally important
and neither is rare.  Discovery is important in astronomy because
we are not sufficiently imaginative to construct the Universe and
its constituents from first principles. Discoveries are needed to
guide us to the choices made by the Universe.  Understanding requires
detailed observations (usually spectroscopy or detailed time series)
and theoretical (and increasingly numerical) analysis.

Take, for instance, the case of  dark matter. The expectation of
dark matter did not come from theoretical considerations. Astronomers
making some of the most basic measurements, in this instance inferring
the mass of galaxies and clusters via rotation curves, were forced
to consider non-luminous matter or dark matter. Supernovae of the
type Ia were studied by astronomers who were curious to know more
about (the then) brightest explosions and  entirely innocent of the
future importance of these explosions for cosmography. We still do
not know what makes Ia supernovae explode but their use as standard
candles led astronomers to a model of an accelerating Universe.
Neither of these great advances were a result of a well laid-out
physics experiment but resulted from astronomers doing routine (and
curiosity driven) research.

In astronomy, in most instances, the importance of a discovery
is not immediately clear. It is because in Astronomy, unlike in Physics,
the inference we can make is limited by not merely the precision
of measurements but also by our ability to ``marginalize out''
unrelated phenomena.  The case of solar neutrinos is an informative
and illustrative example.  The discrepancy between the measured
flux of solar neutrinos and that expected from solar models was
initially attributed to an incomplete understanding of the working
of the Sun.  Gradually, ever increasing and penetrating helio-seismological
observations showed that the solar neutrino puzzle must have a
different explanation. Thanks to concerted experimental efforts --
detailed observations of Solar neutrinos, atmospheric neutrinos and
accelerator neutrinos -- and a sound theoretical framework the case
for neutrino oscillations was established.

A rare counter-example is the discovery of Cosmic Microwave Background
(CMB) radiation.  Unlike the two examples given above the basic
physics of the CMB intensity is very well understood (involving
photons and electrons) and once astronomers obtained 1 part per
million precision in measurements the ensuing progress in our
understanding of the Early Universe has been truly extraordinary.

These two examples lead me to conclude that it is unlikely that we
will see secure and rapid progress in further astronomical investigation
of dark energy.  This is because unlike the solar neutrino problem
we do not have supporting laboratory measurements. Next, the
theoretical foundation of dark energy is murky.  Furthermore, the
several proposed astronomical diagnostics of dark energy (cosmography
and large scale structure) are subject to a host of systematics
measures and as a result will have limited accuracy\footnote{The
ineffective search for dark matter via observations of cosmic
$\gamma$-ray sky illustrates the vast gulf between naive physics-based
aspirations  and astronomical reality.}.

\section{What does this decade hold?}

Astronomy has enjoyed a golden period and all indications are that
this golden phase will continue into this decade and beyond.
Nonetheless, given the financial situation (\S\ref{sec:Background})
it is important to identify those sub-fields where the greatest
progress is expected and ensure that adequate support is provided
for such fields.

In my opinion there are three fields
that will most certainly enjoy great growth during
this period.This is not an exhaustive list and I will leave it
to other proponents to argue for additional fields. The fields I
have in mind\footnote{I note that 
these three fields and experimental gravitational wave astronomy are explicitly 
called out as Discovery Areas by Astro2010.} are Extra-solar planets (\S\ref{sec:ExtraSolarPlanets}),
Astrometry (\S\ref{sec:Astrometry}) and Transients
(\S\ref{sec:Transients}). Owing to my much greater familiarity with
the field the case for transients is more developed.

\section{Extra-solar Planets}
\label{sec:ExtraSolarPlanets}

Extra-solar planets as a field did not exist until 1992. This field
is now sizzling and has an assured future in this decade. A rich
panoply of methodologies (occultation, precision RV, micro-lensing,
high contrast imaging \&\ spectroscopy) is now being pressed by
astronomers to study extra-solar planets.  The costs for ground-based
activities are modest: new high precision spectrographs for large
telescopes, a dedicated network of telescopes for micro-lensing,
radial velocity and occultation (in most instances this could be
reuse of existing telescopes or a collection of inexpensive small
telescopes) and exploiting extreme AO systems that will shortly
come on-line on large telescopes (e.g. GPI on Gemini; P3K+P1640 on
Palomar). The space-ground synergy of Kepler+Keck can be expected
to yield a steady stream of spectacular results.  The end of the
decade should see a clear understanding of planetary architectures
and this will constitute a fundamental and important advance in
astronomy (and with ramifications that go well beyond all of science).

\section{The Era of Astrometry \&\ the importance of highly Multiplexed
Spectroscopy} 
\label{sec:Astrometry}

Astrometry is perhaps the oldest of astronomical methodologies.
However, over the past century, photometry, spectroscopy and adaptive
optics flourished and rose to prominence. Along with many astronomers I see
\textit{Gaia} in making astrometry as powerful as these methodologies.
The combination of \textit{Gaia}, VLBA and laser guide AO-assisted moderate
(e.g. GLAO on MMT; [\textit{C}])
and narrow (e.g. NGAO on Keck) field-of-view astrometry will usher in an era of
ten to hundred micro-arcsecond astrometry.  In tandem, ground based
surveys (UKIDSS, PanSTARRS-1, VST, SkyMapper) will extend 
centi-arcsecond astrometry to fainter magnitudes and NIR bands.
Separately, these surveys will complete the multi-band digital
imaging revolution started by SDSS.  I expect great progress in
this field simply because of the incredible abundance of new data
compared to what exists now.

As dramatically illustrated by SDSS, photometry when accompanied
by spectroscopy has a great multiplicative effect.\footnote{The
opposite is equally true.  Large photometric surveys without
appropriate spectroscopic and related follow up will have limited
impact -- a concern that PR2012 is well advised to ponder  in regard to
LSST.} For this reason, the anticipated revolution from \textit{Gaia},
PS-1 and SkyMapper will only be completed with abundant availability
of spectrographs with massive multiplexing (many thousands of
channels).

The gains will be primarily in stellar and Galactic astronomy
(including ``near field cosmology").  The primary cost is in funding 
astronomers to 
exploit the trove of data that we expect from these missions and
projects. Highly efficient single object spectrographs on existing
medium and large telescopes will help astronomers study unique
objects whereas BigBOSS \& PFS (also on existing telescopes) will
result in a comprehensive study of our Galaxy (elemental abundance,
dynamics, mass distribution).

\section{Transients}
\label{sec:Transients}

A century
ago, the study of variable stars was a major focus of the biggest
astronomical observatories.  Thanks to Moore's law operating not
just for computing but also for transmission of data, for  storage
\textit{and} for optical sensors, astronomers are now able to build,
at relatively low cost, large field-of-view optical cameras and
undertake analysis and rapidly transmit their results for follow-up
observations.  Radio astronomy is on the verge of undergoing a
similar revolution (for similar reasons). Transient object astronomy is
poised to be come a growth field during this decade.

Elsewhere\footnote{The article, originally written for IAU Symposium
285 on Time Domain Astronomy, can be found at
\texttt{http://www.astro.caltech.edu/\textasciitilde srk/PTFOxford.pdf}.
This paper also introduces the planned Zwicky Transient Facility
(ZTF), discussed at the end of this section.} I have noted the great
value of highly focused synoptic surveys.  The Catalina Sky Survey,
a dedicated Near Earth Object survey based on 1-m class telescopes,
discovered and enabled rapid prediction of the entry point of  NEO
2008TC3 [\textit{B}] -- the cheapest sample return mission.

Another cost-effective and focused project and also based on two
aging 1-m class telescopes is the Palomar Transient Factory
(PTF)\footnote{\texttt{http://www.caltech.edu/ptf}}.  In only two
years of operation,  PTF has classified more than 1400 supernovae
(and detected probably three times more candidates). A new class
of luminous supernovae (and ascribed to the deaths of the most
massive stars), the de-lineation of sub-classes of supernovae (some
linked to the coalescence of white dwarfs) and the rapid discovery
of a nearby Ia supernova (within 11hours of the explosion) are some
highlights of this project.

The first experimental detection of gravitational waves from
coalescing neutron star binary will truly constitute a great advance
in both physics and astronomy.  NSF has invested large sums of money
in experimental GW astronomy (pre-LIGO, LIGO, eLIGO and aLIGO).
Subsequent progress in this field would \textit{require} electro-magnetic
localization (if only to set the physical scale based on the redshift
of the host galaxy).  To do so requires not merely the detection
of the electromagnetic counterpart but elimination of a vast fog
of fore- and back-ground transients (e.g. [\textit{D}] and related
papers submitted to Astro-2010).  By 2018 (aLIGO era) we will have
sufficient assets (PS-1, ZTF, SkyMapper, DES, ODI, EVLA and rapid
spectroscopy on large telescopes) to realistically pursue EM
counterparts of nearby events ($\lesssim 150\,$Mpc).

Separately, the current suite of synoptic surveys is very well
suited to explore the phase space of transient searches (for which
follow-up spectroscopy is essential; the current stable of 4-m and
8-m telescopes is adequate for this task). Future larger facilities
will be only useful for those transients which are 
intrinsically faint or apparently fainter (great cosmological distance).  Synoptic surveys done on larger telescopes
will provide better photometric precision and LSST is  better suited
for subtle phenomenology in variable stars.

So great are the promises of this field that along my colleagues
and I are working towards an integrated \textit{facility} with a
triad of dedicated telescopes (Zwicky Transient Facility; ZTF$^{\tiny
4}$) equipped as follows: a very large field-of-view (40 square
degrees) imager, a low resolution classification spectrometer\footnote{See
position paper submitted to PR2012 by N. Konidaris. The spectrometer 
mounted on the 84-inch KPNO
telescope could constitute the first element  of The National TOO
Telescope System.} and a robotic AO photometric
machine\footnote{C. Baranec ({\it ibid}) describes the demonstration of a robotic LGS-AO
on the Palomar 60-inch telescope.}. ZTF aims to explore the 
phase space of short duration transients with a
bread-n-butter focus of studying rise-time phenomenology.

\section{An optimal balance}
\label{sec:OptimalBalance}

Here, I argue that progress in astronomy rests on making discoveries
and obtaining an appropriate depth of understanding. Discoveries
usually arise from large surveys (e.g. the Hulse-Taylor pulsar, a
landmark object; high redshift quasars from SDSS \&\ UKIDSS; unveiling an
entirely new spectroscopic class (Y \& L) resulting from  WISE
and 2MASS surveys or from dedicated programs almost always on smaller
telescopes (e.g. the first brown dwarf, the spectral class T  and the first planet around
a normal star) or from development of new techniques or methodology
(e.g.  cosmic X-ray astronomy).

In contrast, depth requires detailed study and almost always comes
from large telescopes (e.g. the shrinkage of the orbit of the
Hulse-Taylor pulsar from Arecibo timing observations; the onset of
the Gunn-Peterson effect from sensitive spectroscopic observations
of SDSS quasars undertaken at Keck Observatory) followed by or
preceded by modeling and theoretical studies (which, in many cases,
requires laboratory studies).

Accepting the equal role of Discovery and Depth our resource
allocation should be close to equipartition. Unfortunately, in
practice, influential astronomers start large projects based purely
on aspirations (``bigger is better'') and with virtually no
cost-benefit analysis.  Furthermore, the annual operation cost of
modern facilities has increased\footnote{There are sound reasons
why there has been an increase from the traditional few percent to
10\%.} and is roughly 10\% of the capital cost\footnote{In the
annual operating cost I  include all related expenditures and in
particular the cost of new instrument (appropriately amortized)
that help keep the flagship enterprise at the forefront.}. As a
result we now talk of ``life-cycle'' costs (the sum of the capital
\textit{and} operating costs over the lifetime of the facility).
The longevity of astronomical facilities means that the \textit{Opportunity
Cost}\footnote{Well known to businesses who have to balance their
financial books on timescales much shorter than that for educational
institutions or government aided research organizations.}[\textit{A}]
of flagship projects now extends over nearly a professional lifetime
(a decade for construction and two for operations).

In times of easy finance one does not have to think hard about
undertaking cost-benefit analysis.  Astronomy in the US (and to
some degree in Europe as well) literally enjoyed a bubble with a
rapid  increase of funding over the past three decades.  This bubble
of funding coincided with a genuine bubble of ideas.  The financial
bubble has now deflated and we are suffering withdrawal symptoms,
first in space\footnote{All points raised here are already fully
applicable to NASA investments in Astronomy -- thereby providing
some support to the conclusions presented here.} -- and now in
ground-based astronomy.

\section{Recommendations}
\label{sec:Recommendations}

In summary, as we enter this decade and assuming that the funding
for AST
remains constant\footnote{All indications are that the US economy
will, at best, grow modestly through this decade and more likely
stagnate for the entire decade.}, my advice to PR2012 is to preserve
our strength in discoveries, redeploy existing assets creatively
and selectively fund sub-fields with the largest growth potential.
These suggestions satisfy the true boundary condition for PR2012
-- the financial boundary condition whilst ensuring astronomical
returns.

Under the rubric of mid-scale projects, groups could be funded to
build massively multiplexed spectrographs on existing telescopes
(\S\ref{sec:Astrometry}); build an integrated network of (re-purposed)
telescopes  to comprehensively explore the transient sky to 22 mag
(\S\ref{sec:Transients});  and undertake innovative but affordable
projects in extra-solar planets (\S\ref{sec:ExtraSolarPlanets}).
PI grants to take advantage of ground-space synergy (\textit{Gaia}
and Explorer programs) and astrometric surveys are needed.  I make
a note that this mix of projects is very well suited to the ecology
of the American astronomical society which drives its strength from
University-based research.  Mid-scale projects train more than pure
observers and data analysts. They  train young astronomers in hardware, software
and methodologies and  enrich not just mere academia but also
industries and research laboratories.

Next, a critical re-examination of  the cost-benefit analysis for
both approved and near future flagship enterprises is essential.\footnote{I realize that suggestions
made in this section are likely to be seen as inconsistent with the
boundary conditions for PR2012 laid down by NSF, namely, recommendations
to PR2012 cannot substantially alter the recommendations nor change
priorities of the last NRC Astronomy Review (Astro2010).  Reworking
at the margins is unlikely to make a major difference to the future
of American astronomy in this decade. If the financial boundary
conditions are what are claimed to be then the current situation
calls for bold leadership and tough decisions and not merely the
proverbial rearrangement of the chairs in the dining hall of the Costa Concordia
as it approached the Isola del Giglio.}
This decade calls for sharp analysis, bold leadership and creative solutions along
the spirit of ``Less is More'' rather than the ``More is More''
mantra of the go-go decades. The alternative will certainly lead to the
decline of US leadership in global astronomy.

Separately, US astronomers should recognize that the steering of vast
funding for Dark Energy will have (and is having) a deleterious
effect on astronomical  exploration  of the sky
(cf. [\textit{E}]). PR2012 should encourage the Department of Energy
(DOE) to take over the  \textit{entire} program on Dark Energy 
(including  the chimera nourished and forged by the heat generated by Astro2010 deliberations). 
American astronomers, relieved of  this crushing financial burden, can, over this decade, 
continue to do what they do best -- make headlines with real discoveries 
and  increase our understanding of the
incredibly rich heavenly sky through meaningful observations.


{\small
\par\noindent [\textit{A}] Bastiat, F. 1848, ``What is seen and what is not seen''
\par\noindent [\textit{B}] Boattini, A. et al. 2009, DPS 41, \#9.02
\par\noindent [\textit{C}] Hart, M. et al. 2010, Nature 466, pp. 727 
\par\noindent [\textit{D}] Phinney, E. S. 2009, arXiv 0903.0098
\par\noindent [\textit{E}] White, S. 2007, Rep Prog Phys 70, pp. 883
 }
\end{document}